\documentclass[showpacs,amsmath,amssymb,prb,superscriptaddress,,reprint]{revtex4}
\usepackage{graphicx}
\usepackage{multirow}
\usepackage{textcomp}
\usepackage{color} 

\usepackage[colorlinks,plainpages=false,linkcolor=blue,urlcolor=blue,citecolor=blue,pdfpagemode=UseNone,pdfstartview=FitBH]{hyperref}
\newcommand{\picene}{C$_{22}$H$_{14}$}

\begin{document}
\DeclareGraphicsExtensions{.eps, .jpg, .pdf}

\title{Vibrational spectrum of solid picene (\picene)}

\author{B. Joseph} 
\affiliation{Dipartimento di Fisica, Universit\`{a} ~di Roma Sapienza, 
P.le Aldo Moro 2, 00185 Roma, Italy }
\author{L. Boeri}
\affiliation{Max Planck Institute for Solid State Research, Heisenbergstrasse 1, D-70569, Stuttgart, Germany}
\author{L. Malavasi} 
\affiliation{Dipartimento di Chimica, Universit\`{a}  di Pavia, Via Taramelli, 27100 Pavia, Italy}
\author{F. Capitani} 
\affiliation{Dipartimento di Fisica, Universit\`{a} ~di Roma Sapienza, 
P.le Aldo Moro 2, 00185 Roma, Italy }
\author{G.A. Artioli} 
\affiliation{Dipartimento di Chimica, Universit\`{a}  di Pavia, Via Taramelli, 27100 Pavia, Italy}
\author{S. Protti}
\affiliation{Dipartimento di Chimica, Universit\`{a}  di Pavia, Via Taramelli, 27100 Pavia, Italy}
\author{M. Fagnoni}
\affiliation{Dipartimento di Chimica, Universit\`{a}  di Pavia, Via Taramelli, 27100 Pavia, Italy}
\author{A. Albini}
\affiliation{Dipartimento di Chimica, Universit\`{a}  di Pavia, Via Taramelli, 27100 Pavia, Italy}
\author{C. Marini}
\affiliation{European Synchrotron Radiation Facility, BP 220, 38043 Grenoble 
Cedex, France}
\author{L.~Baldassarre\footnote{Present address:
Center for Life NanoScience@LaSapienza, Istituto Italiano di Tecnologia, Viale Regina Elena 295, Roma, Italy}} 
\affiliation{Sincrotrone Trieste, S.C.p.A., Area Science Park, I-34012, Basovizza, Trieste, Italy}
\author{A. Perucchi} 
\affiliation{Sincrotrone Trieste, S.C.p.A., Area Science Park, I-34012, Basovizza, Trieste, Italy}
\author{S. Lupi} 
\affiliation{CNR-IOM and Dipartimento di Fisica, Universit\`{a} di Roma Sapienza, 
P.le Aldo Moro 2, 00185 Roma, Italy}
\author{P. Postorino} 
\affiliation{CNR-IOM and Dipartimento di Fisica, Universit\`{a} di Roma Sapienza, P.le Aldo Moro 2, 00185 Roma, Italy}
\author{P. Dore}\email[corresponding author: ]{Paolo.Dore@roma1.infn.it}
\affiliation{CNR-SPIN and Dipartimento di Fisica, Universit\`{a} di Roma Sapienza, P.le Aldo Moro 2, 00185 Roma, Italy }


\begin{abstract}

\bigskip

Recently, Mitsuhashi {\it et al}  have observed superconductivity with transition
temperature up to 18 K in potassium doped picene (\picene), a polycyclic
aromatic hydrocarbon compound [{\it Nature} \href{http://dx.doi.org/10.1038/nature08859}{ 464 (2010) 76}]. Theoretical
analysis indicate the importance of electron-phonon coupling in the
superconducting mechanisms of these systems, with different
emphasis on inter- and intra-molecular vibrations, depending
on the approximations used. Here we present a combined experimental and ab-initio 
study of the Raman and infrared spectrum of undoped solid picene, which allows us to unanbiguously assign the vibrational modes. This combined study enables the identification of the modes which couple strongly to electrons and hence can play an important role in the superconducting properties of the doped samples.
 
\bigskip

Journal reference : {\it Journal of Physics: Condensed Matter} \href{http://dx.doi.org/10.1088/0953-8984/24/25/252203}{ 24 (2012) 252203}

\end{abstract}


\maketitle

\newpage
Recently, superconductivity has been observed in potassium doped picene, with transition temperatures T$_c$ of 7-18 K \cite{Mitsuhash2010}. 
This was the first report on "high-$T_c$"  superconductivity (SC) in an aromatic compound.
Picene (C$_{22}$H$_{14}$) is an alternant polycyclic aromatic hydrocarbon (PAH), {\em i.e.}
a planar aromatic molecule, formed by juxtaposed benzene rings.
Picene comprises five rings, arranged in a zig-zag fashion.
After this initial report, 
SC was also reported in phenanthrene (three rings)~\cite {Wang2011_SC-phenanthrene}, dibenzpentacene (five + two rings)~\cite{Xue2011_SC-dipicene} and coronene (six rings)~\cite{Mitamura2011_perpestive-picene}  
upon doping with  alkali or alkalki-earth metals.  
SC was then also observed in picene and phenanthrene 
doped with rare-earths \cite{Mitamura2011_perpestive-picene,Wang2011_SC-phenanthrene2,Wang2012_SC-phenanthrene3},
indicating that the PAHs  most likely form a completely new class of superconductors, with highly {\em tunable}
electronic properties.

Due to their 2-dimensional nature, PAHs are naturally related to other {\em layered} superconductors, such as MgB$_2$~\cite{Nagamatsu2001_MgB2},
or the newly-proposed graphane~\cite{Savini2010_SC-graphane}.
However, in contrast to all these systems, which are conventional electron-phonon ($ep$) superconductors, the nature and origin of
SC in PAH are still debated. Theoretical studies have focused either on electron-electron ($ee$)~\cite{Giovannetti2011_electronicStr,Kim2011_electronicStr} 
or $ep$ interaction~\cite{Casula2011_ele-ph,Subedi2011_ele-ph,Kato2011_ele-ph,Sato2012_ele-ph},
which are both sizable in these $\pi$-bonded systems.
However, estimates of strength and spectral distribution of the $ep$ interaction 
differ depending on the approximation used for doping. The main
question is to estabilish 
how much the $ep$ coupling in an isolated molecule picture is representative of coupling in the actual doped compound~\cite{Casula2011_ele-ph,Subedi2011_ele-ph,Kato2011_ele-ph}.
A definite answer on this point could only come from a detailed experimental comparison of the phonon spectra and lifetimes of the doped and undoped samples. This has not been 
possible so far, because the vibrational spectra are extremely rich, and the samples are  poorly characterized.
For the doped superconducting samples in ref.~\cite{Mitsuhash2010}, not only the detailed crystal structure, but also the precise doping level are unkwown. 
For pure picene, X-ray diffraction profiles have been refined several years ago ~\cite{De1985_XRD-picene}, but, owing to the lack of high-purity samples, there were no serious attempts to a complete characterization of the vibrational spectrum, except for a recent study of the optical phonons~\cite{Girlando2012_IR_Ram_picene}, which appeared in the literature while we were finalizing the present manuscript. 

\begin{figure}
\includegraphics[width=7.5cm]{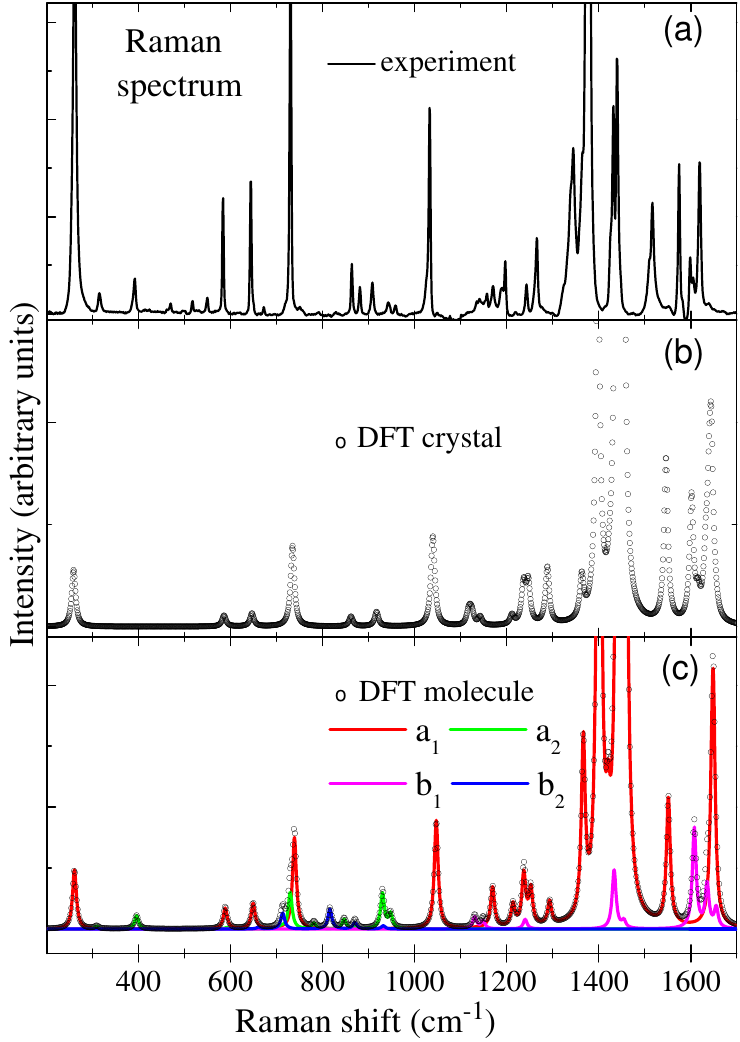}
\caption{(Color online) Comparison of the experimental Raman spectrum of crystalline picene (a) with the DFT spectrum of crystal (b) and  relaxed molecule (c). For the molecule, the spectral components of different symmetry ($a_1$, $b_1$, $a_2$, $b_2$) are shown separately (see text for details).  The computed DFT stick spectra are convoluted with a Lorenzian profile with 5 cm$^{-1}$ linewidth to ease comparison with experiment.}
\label{fig:1}
\end{figure}


In this paper, we present a combined experimental and theoretical
study~\cite{ddicastro,cmarini} of the vibrational spectrum of crystalline picene, based on accurate  Raman and infrared measurements at room temperature on high quality samples. 
Density functional perturbation theory (DFPT~\cite{Baroni2001_DFPT}) calculations were also performed, which allowed us to obtain a detailed assignment of vibrational modes. This study is propaedeutical to further studies on the high-pressure behaviour of doped and undoped samples, which could lead to significant insights into the importance of intermolecular coupling of vibrational modes and electrons in hydrocarbons, and on the  reliability of DFT in organic compounds~\cite{blase}.  Besides superconductivity, our study 
is of interest also for potential applications in organic electronics~\cite{Mitamura2011_perpestive-picene,Okamoto2008_FET-picene}.

\begin{figure}
\includegraphics[width=7.5cm]{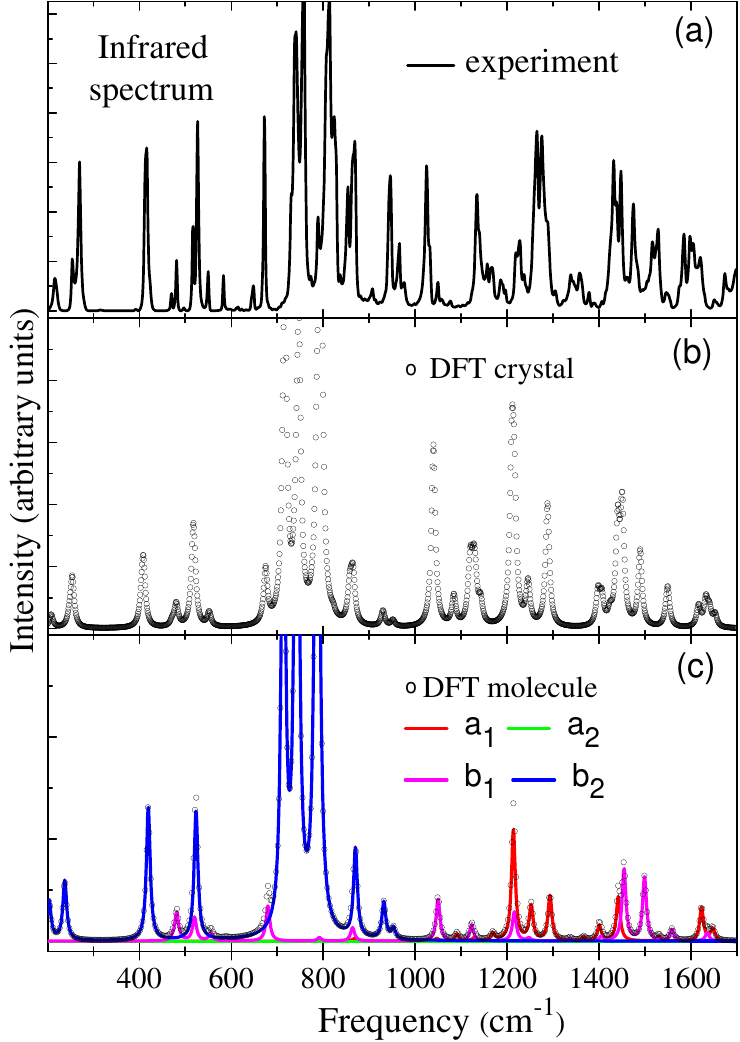}
\caption{(Color online) Same as Fig. \ref{fig:1}, for the IR spectra.}
\label{fig:2}
\end{figure}

Solid picene was prepared by a new optimized synthesis route which permits to obtain bulk quantities (hundreds of milligrams in one shot) of pure polycrystalline picene powder~\cite{Malavasi2011_picene-prep}. Refinements of the X-ray diffraction spectrum of our sample yielded results identical to those previously  reported~\cite{De1985_XRD-picene}.  
We performed Raman measurements by means of a confocal-microscope Raman spectrometer, equipped with a He-Ne laser ($\lambda$=632.8 nm), a 1800 lines-mm$^{-1}$ grating monochromator, and a charge-coupled-device (CCD) detector. Raman measurements were carried out on the powder samples in the backscattering geometry, and a notch filter was used to reject the elastic contribution.  Using a few micron-sized laser spot on the sample, accurate Raman spectra were collected with a spectral resolution better than 3 cm$^{-1}$. A typical Raman spectrum is reported in Fig. \ref{fig:1}(a) over the frequency range 200-1700 cm$^{-1}$. Note that intensity and width of the highest and sharpest peaks may be affected by the instrumental spectral resolution. For infrared (IR) measurements, we employed a thin pellet (of thickness around 50 micron) prepared out of pure picene powder. IR trasmittance
measurements were carried out at the SISSI infrared beamline of the ELETTRA synchrotron (Trieste, Italy) \cite{Lupi2007_SISSI}.   
Using a Bruker IFS 66v interferometer equipped with both mylar beam splitters coupled to a liquid He bolometer for the far-IR region and a KBr beam splitter coupled to a mercury-cadmium-telluride (MCT) detector for the mid-IR region, accurate IR spectra with a spectral resolution close to 2 cm$^{-1}$ were collected. 
A typical IR spectrum is reported in Fig. \ref{fig:2}(a) over the frequency range 200-1700 cm$^{-1}$. We notice that the most intense absorption lines result from vanishingly small transmitted intensities. These lines can therefore be affected by rather large uncertainties. Nevertheless, a large number of low intensity absorption peaks are determined with a good accuracy. 

The Raman and IR spectra of picene in its crystalline and molecular forms have been calculated using DFT~\cite{Baroni2001_DFPT,Lazzeri2003_Raman_DFT,qe}.
The electron-ion interaction was described using LDA norm-conserving pseudopotentials~\cite{psps};
electronic wavefunctions were expanded on a plane-wave basis, with a 80 Ryd cutoff.
A $2^3$ uniform grid  for $\mathbf{k}$-space integration in the solid allowed an accuracy of $\pm$ 5 cm$^{-1}$ on the calculated frequencies. 
For the crystal, we employed the experimental unit cell, and relaxed atomic positions. The resulting Raman and IR spectra are reported in Figs. \ref{fig:1}(b) and \ref{fig:2}(b), respectively.  For the molecule, we employed periodical repetitions of a 24 $\times$ 10 $\times$ 5 \AA$ $ supercell, containing a single picene molecule,
with the $z$ axis orthogonal to the $ab$ plane, and the long molecular axis along the $x$ axis. The cut-off energy was rescaled accordingly.
We performed calculations both for an ideal geometry, and for a relaxed geometry in which we optimized the internal coordinates
subject to the constraint of symmetry. The resulting Raman and IR spectra are reported in Figs. \ref{fig:1}(c) and \ref{fig:2}(c), respectively. Note that the results we report  in these figures are for the relaxed molecular geometry, but the difference
between the ideal and relaxed molecule spectra is never larger than $\pm$ 10 cm$^{-1}$.

\begin{figure}
\includegraphics[width=7.5cm]{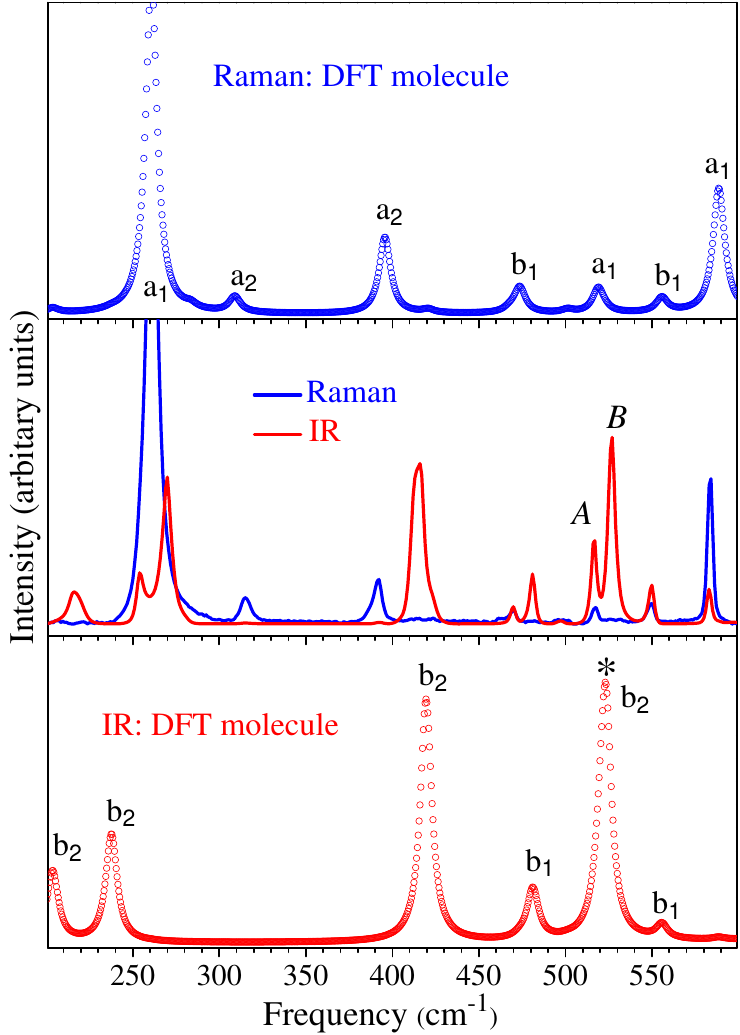}
\caption{(Color online) Comparison between the measured IR and Raman spectra (middle panel) and DFT Raman (upper panel) and IR (lower panel) spectra in the range 200 - 600 cm$^{-1}$. The computed DFT stick spectra are convoluted with a Lorenzian profile with 10 cm$^{-1}$ linewidth to ease comparison with experiment. Similar comparisons at higher frequencies are given in the supplementary information file. In the IR DFT spectrum of the molecule (lower panel), the $b_2$ peak at 523 cm$^{-1}$ marked by $*$  splits into $A$ and $B$ components (see text) in the measured spectrum (middle panel).}
\label{fig:3}
\end{figure}

In comparing the measured and the calculated spectra, it is convienent to recall that
the calculated partial phonon density of states (not shown) indicates
that modes up to 1000 cm$^{-1}$ involve both out-of-plane and in-plane
vibrations of carbon and hydrogen; above 1000 cm$^{-1}$, the vibrations are only
in-plane, with high-energy vibrations above 3000 cm$^{-1}$ involving only H atoms.~\cite{Subedi2011_ele-ph}. 
The unit cell of crystalline picene is monoclinic (space group P2$_1$) and
contains two C$_{22}$H$_{14}$ molecules, arranged in a herringbone fashion and inclining slightly to the $ab$ plane. This yields 216 vibrational modes, many of which are both IR and Raman active.
The comparison of the upper and middle panels of the two figures gives an idea of the accuracy of DFT calculations: DFT captures many of the main features of the experimental spectra, including the positions, intensities and splitting of the main peaks, with a few exceptions. 
The lower panel of the two figures shows the equivalent spectra, calculated
for the picene molecule; the symbols show the total intensities, while
colored lines refer to the four possible symmetries for the vibrational modes.
The picene molecule has C$_{2v}$ symmetry; of the 102 optical modes,
those with  $a_1$ (35), $b_2$ (16), $b_1$ (34) symmetry are both Raman and IR active, while the $a_2$ (17) are only Raman-active~\cite{table:web}. A full list of peak frequency, symmetry, optical activity and the normalized intensity of the vibrational modes is given in the supplementary information file. 
It is worth noting that, for an isolated molecule, the Raman reponse is dominated by the (totally symmetric) a$_1$ modes (Fig. \ref{fig:1}(c)), the IR reponse by the b$_2$ modes (Fig. \ref{fig:2}(c)).  

From a comparison between the calculated crystal and molecule spectra (see Figs. \ref{fig:1} and \ref{fig:2}), it can be observed that the molecular modes dominate the optical response.  Picene 
is thus a typical molecular crystal, in which the optical response is determined by the selection rules due to the point group of the molecule, on which the site group of the crystal acts as a perturbation~\cite{Rousseau1981_RamanNormalMode}.

\begingroup
\begin{table}
\begin{center}
\caption{\label{tab:table1} Assignment of the most intense vibrational modes
observed. For each mode, the frequency, symmetry and optical activity
calculated in DFT are reported together with the corresponding IR and/or
Raman frequency. All frequencies are in cm$^{-1}$. The $a_1$ modes expected to be 
most involved in the $ep$ coupling \cite{Kato2011_ele-ph} are marked by $*$.
A complete list of frequencies and intensities of all the modes calculated in DFT
is provided in the supplementary information file. }
{\small

\begin{tabular}[t]{||c|c|c||c|c||}
\hline
Frequency  &      & 				& Frequency & Frequency \\
 DFT     & Mode & Activity   &  IR  &  Raman    \\
\hline
203.8      &b$_2$ &  I+R     & 217       &       \\
237.8      &b$_2$ &  I+R     & 254 , 270   &       \\
260.8      & a$_1$&	I+R     & 	        &262*    \\
309.0      &a$_2$	&   R      & --        &314    \\
395.8      &a$_2$ &	R       & --        &391    \\
419.5      &b$_2$	&I+R       &415        &       \\
481.2      &b$_1$ &I+R       &481        &       \\
523.3      &b$_2$ &	I+R     &517 , 527  &517    \\
555.8	     &b$_1$ &	I+R     &550	     &550    \\
588.5      &a$_1$ &	I+R     &584	     &584    \\
649.2	     &a$_1$ &	I+R     &   	     &644    \\
679.8      &b$_1$ &	I+R     &672		  &   	\\
713.4		  &b$_2$	&	I+R	  &740 , 756  &		\\
738.9		  &a$_1$	&	I+R	  &			   &730*	\\
742.5		  &b$_2$	&	I+R	   &810  	   &		\\
787.0		  &b$_2$	&	I+R		&853 , 865  &		\\
867.3		  &a$_1$	&	I+R		&			  &864	\\
870.2		  &b$_2$	&	I+R		&		  &		\\
930.2		  &a$_2$	&	 R		   & --		  &908	\\
932.1		  &b$_2$	&	I+R		&947		  &		\\
1047.1	&a$_1$	&	I+R		&			  &1032	\\
1050.1	&b$_1$	&	I+R		&1025		  &		\\
1122.9	&b$_1$	&	I+R		&1134		  &		\\
1213.9	&a$_1$	&	I+R		&1227		  &		\\
1252.6	&a$_1$	&	I+R		&1265		  &1265	\\
1293.5	&a$_1$	&	I+R		&1276		  &		\\
1366.7	&a$_1$	&	I+R		&			&1345	\\
1400.8	&a$_1$	&	I+R		&   		&1377*	\\
1442.1	&a$_1$	&	I+R		&1433		&1433	\\
1454.7	&b$_1$	&	I+R		&1447		&		\\
1456.2	&a$_1$	&	I+R		&			&1440	\\
1499.3	&b$_1$	&	I+R		&1475		&		\\
1551.5	&a$_1$	&	I+R		&   		&1516*	\\
1607.9	&b$_1$	&	I+R		&			&1574	\\		
1649.2	&a$_1$	&	I+R		&  		&1620	\\
\hline
\end{tabular}}
\end{center}
\end{table}

Given the overall agreement between the spectra calculated for the molecule and the crystal, we have used the molecular symmetry assignments to label the vibrational modes observed in the solid. Fig.~\ref{fig:3} shows a detailed comparison between the experimental (Raman and IR) and the calculated  (molecule) spectra in the energy range between 200 and 600 cm$^{-1}$. Similar comparisons at higher frequencies are given in the supplementary information file. In the same file we also report results obtained in the frequency range of the C-H stretching (3000-3250 cm$^{-1}$), which is not of direct interest in the present work. Exploiting this kind of plots and of comparisons, we have been able to make a reliable assignments of most of the observed vibrational peaks. 

In Table 1, we provide the list of the most intense peaks which can be assigned without ambiguity. Our results extend a recent study \cite{Girlando2012_IR_Ram_picene}, which contains a partial assignment of the vibrational modes of picene, based on optical measurements and calculations \cite{note:sym}. Note that in correspondence to some of the DFT lines computed for the molecule, a doublet is observed; this effect can be attributed to the Davydov splitting of two modes corresponding to the same molecular vibration \cite{Girlando2012_IR_Ram_picene}. For example, in Fig. \ref{fig:3} (see lower panel) the $b_2$ peak at 523 cm$^{-1}$ marked by $*$ splits into two components in the measured spectrum (see middle panel). In the table, two values corresponding to the same DFT mode indicate the observed splitting. 

The precise assignment of the phonon frequencies is particularly important, in connection with the problem of $ep$ coupling, which in PAHs is usually 
calculated taking into account only vibronic contributions (see for example Refs.~\cite{Kato2011_ele-ph,Sato2012_ele-ph}).
For picene doped with mono-, bi- 
and tri-anions, as well as with mono-cations, the coupling is non-zero only for
modes with $a_1$ symmetry. According to Ref.~\cite{Kato2011_ele-ph}, in the case of 
bi-anion doping, the modes most involved in the 
$ep$ coupling are at 253, 724, 1380, 1516 cm$^{-1}$. The corresponding values we observe
for pure picene are 262, 730, 1377 and 1516 cm$^{-1}$ (marked by $*$ in Table 1).
A recent calculation by {\em Casula et al.} for K$_3$ picene~\cite{Casula2011_ele-ph}
has shown that this picture is strongly modified by the presence of the potassium, which mediates a strong intermolecular interaction. 
This result is strongly dependent on the geometrical distribution and concentration
of the dopants, and is not representative of {\em pure} picene.

\begin{figure}
\includegraphics[width=7.5cm]{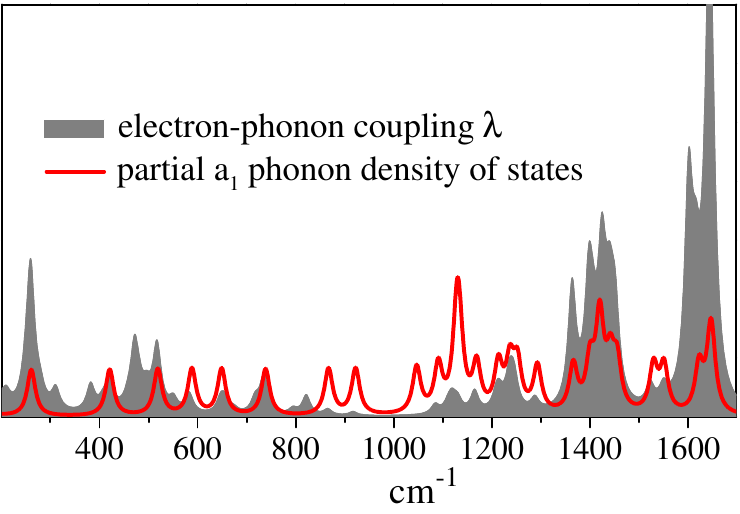}
\caption{(Color online) Distribution of the electron-phonon
coupling at the $\Gamma$ point, calculated in the rigid-band approximation in Ref.~\cite{Subedi2011_ele-ph} and partial $a_1$ phonon density of states. For details of the calculation see \cite{eph:details}.}
\label{fig:eph}
\end{figure}

In Fig.~\ref{fig:eph} we compare the spectral distribution of the $ep$ coupling $\lambda$, calculated in Ref.~\cite{Subedi2011_ele-ph} using the rigid band approximation, with the partial $a_1$ phonon density of states to estimate the effect of intermolecular interactions of {\em pure} crystalline picene \cite{eph:details}.
The figure shows that the $ep$ coupling response is indeed dominated by the $a_1$
modes, but there are additional contributions clearly coming from modes with different
symmetries. Consistently with what we observed in the IR and Raman spectra, the
figure shows that intermolecular interactions in {\em undoped} picene are quite small;
this picture could be modified by doping and/or pressure, which could both
induce a stronger intermolecular coupling, due to hybridization and/or
increased overlap between the molecules.

In conclusion, we have presented high-quality optical data and first-principles 
calculations for the vibrational spectrum of solid picene. The comparison between experimental and theoretical results enabled us to completely characterize the 
spectrum, and assign most of the observed vibrational modes. Comparing the theoretical 
calculations for Raman, IR and electron-phonon spectral function, we have identified 
picene as solid, with a small intermolecular interaction.
The extensive characterization obtained in this study is fundamental for further 
studies on the effect of pressure and/or doping on bonding and intermolecular 
interactions in doped polycyclic aromatic hydrocarbons.\\

\bigskip

\section*{Supplementary information}

\newpage
\subsection*{DFT phonon modes in the frequency range 200-1700 cm$^{-1}$. For each mode, calculated DFT frequency, symmetry, optical activity, and intensity of the IR and/or Raman frequency are reported. Intensities are normalized to a maximum value of 100.00. }

\begin{figure}
\includegraphics[width=17.0cm]{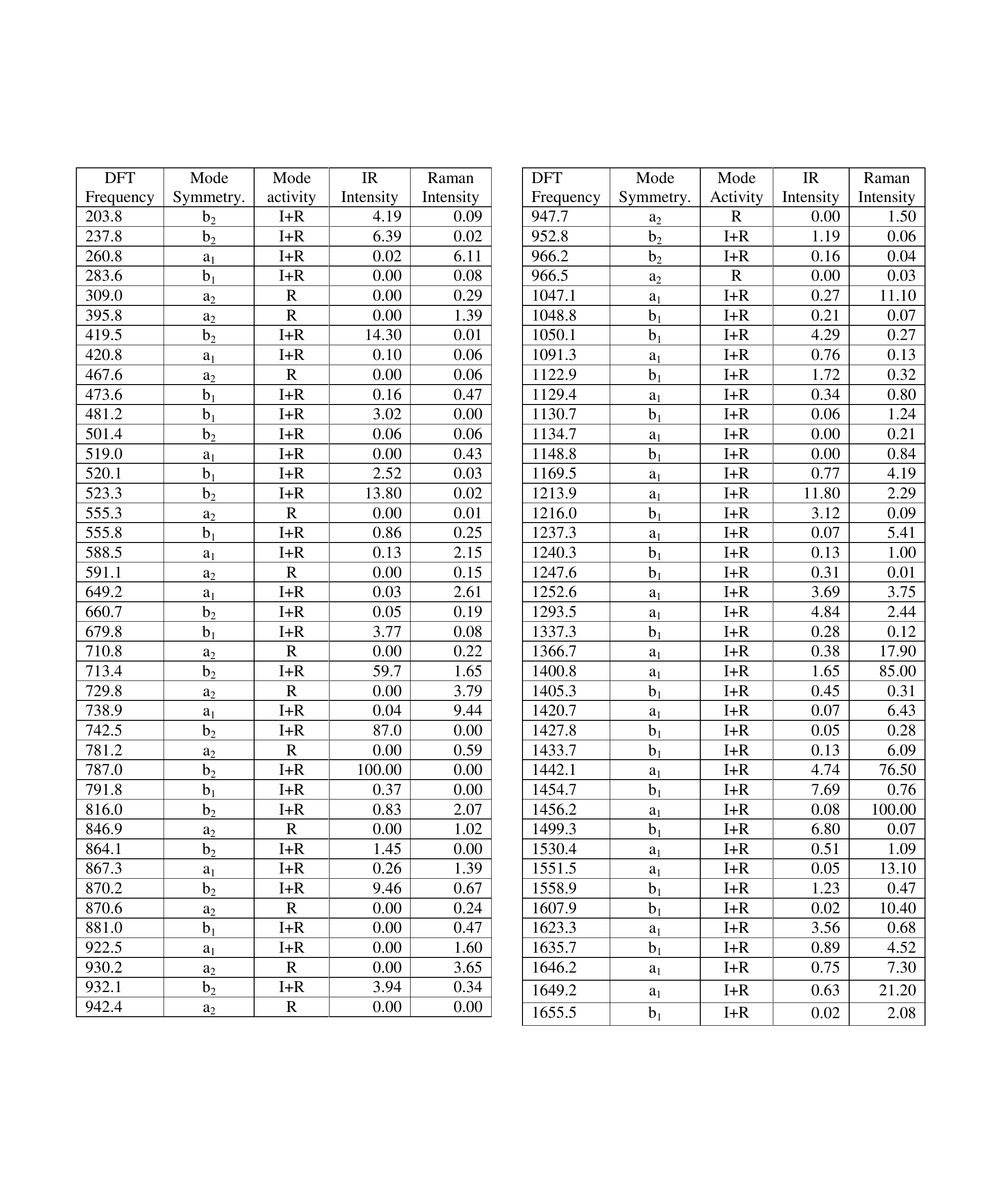}
\label{fig-SI-1}
\end{figure}

\newpage

\begin{figure}
\includegraphics[width=7.0cm]{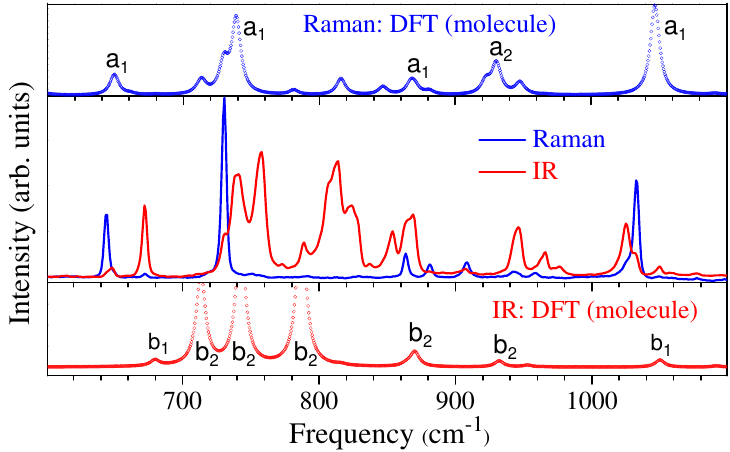}
\caption{Comparison between the measured IR and Raman spectra (middle panel), DFT Raman (upper panel), and IR (lower panel) spectra in the range 600 - 1100 cm$^{-1}$. The computed DFT stick spectra are convoluted with a Lorenzian profile with 10 cm$^{-1}$ linewidth to ease comparison with experiment.}
\end{figure}

\begin{figure}
\includegraphics[width=7.0cm]{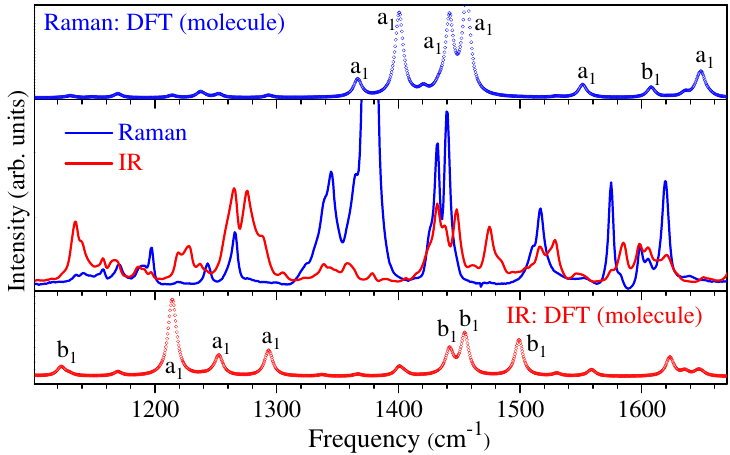}\nonumber
\caption{Comparison between the measured IR and Raman spectra (middle panel), DFT Raman (upper panel), and IR (lower panel) spectra in the range 1100 - 1700 cm$^{-1}$. The computed DFT stick spectra are convoluted with a Lorenzian profile with 10 cm$^{-1}$ linewidth to ease comparison with experiment.}
\end{figure}

\begin{figure}
\includegraphics[width=6.5cm]{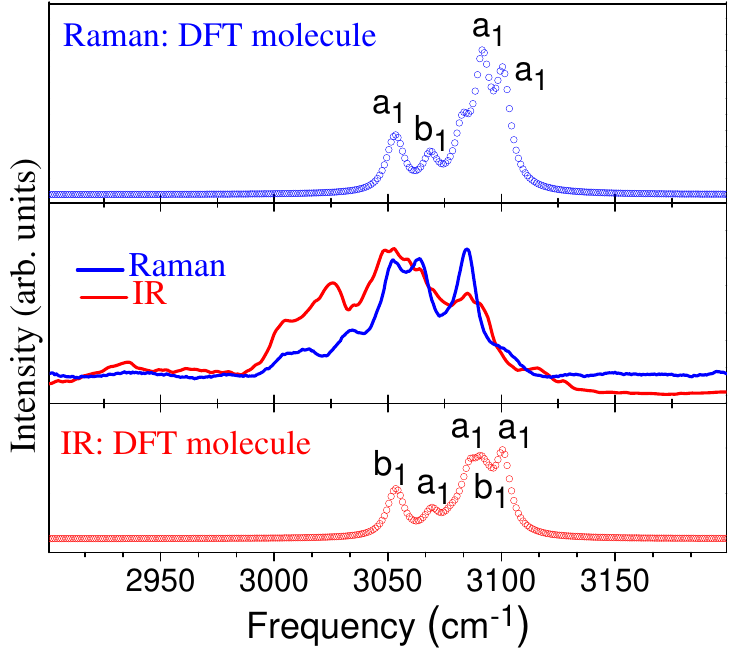}
\caption{Comparison between the measured IR and Raman spectra (middle panel), DFT Raman (upper panel), and IR (lower panel) spectra in the frequency range of the C-H stretching (3000-3250 cm$^{-1}$) which is not of direct interest in the present paper. Notice that the measured Raman and IR spectra have been normalized to the same maximum. Like in other frequency ranges, the computed DFT stick spectra are convoluted with a Lorenzian profile with 10 cm$^{-1}$ linewidth to ease comparison with experiment.}
\end{figure}

\end{document}